# Nanoscopy Reveals Metallic Black Phosphorus

Yohannes Abate[1,2,*], Sampath Gamage[2], Li Zhen[3], Stephen B. Cronin[3], Han Wang[3], Viktoriia Babicheva[1,2], Mohammad H. Javani[1,2], Mark I. Stockman[1,2,*]

1. Center for Nano-Optics (CeNO), Georgia State University, Atlanta, Georgia 30303, USA
2. Department of Physics and Astronomy, Georgia State University, Atlanta, Georgia 30303, USA
3. Viterbi School of Engineering, University of Southern California, Los Angeles, CA 90089, USA

* Corresponding authors, e-mails: yabate@gsu.edu, mstockman@gsu.edu

**Layered and two-dimensional (2D) materials[1] such as graphene[2,3], boron nitride[1,4], transition metal dichalcogenides[1,5-8] (TMDCs), and black phosphorus (BP)[1,9-13] have intriguing fundamental physical properties and bear promise of numerous important applications in electronics and optics. Of them, BP[11,12,14] is a novel 2D material that has been theoretically predicted to acquire plasmonic behavior for frequencies below ~0.4 eV when highly doped. The electronic properties of BP are unique due to an anisotropic structure, which could strongly influence collective electronic excitations. Advantages of BP as a material for nanoelectronics and nanooptics are due to the fact that, in contrast to metals, the free carrier density in it can be dynamically controlled by electrostatic gating, which has been demonstrated by its use in field-effect transistors.[9] Despite all the interest that BP attracts, near-field and plasmonic properties of BP have not yet been investigated experimentally. Here we report the first observation of nanoscopic near-field properties of BP. We have discovered near field patterns of outside bright fringes and high surface polarizability of nanofilm BP consistent with its surface-metallic, plasmonic behavior at mid-infrared (mid-IR) frequencies. This behavior is highly frequency-dispersive, disappearing above frequency, $\omega$ =1070 cm$^{-1}$, which allowed us to estimate the plasma frequency and carrier density. We have also observed similar behavior in other 2D semiconductors such as TMDCs but not in 2D insulators such as boron nitride. This new phenomenon is attributed to surface charging of the semiconductor nanofilms. This discovery opens up a new field of research and potential applications in nanoplasmonics and optoelectronics.**

In contrast to graphene, a semimetal, BP is a semiconductor whose band gap, depending on the number of atomic layers, is narrower than that of TMDCs. In contrast to multilayer TMDCs that have indirect band gap, BP exhibits direct band gap properties regardless of its thickness, which is a significant advantage for optoelectronic applications. Also, BP has a rather high carrier mobility (~$10^4$ cm$^2$ V s$^{−1}$),[15,16] high on/off ratios in transistors, strong excitonic effects and strongly anisotropic optical, electrical and thermal conductance,[16] which is due to its puckered crystalline structure. These physical properties can be manipulated by altering its layer thickness[17,18], stacking order,[19] applied strain force [20] and external electric field [21], offering tremendous advantages for using BP in numerous potential applications in optics and electronics. The anisotropic thin-film BP carrier density can be manipulated by changing doping levels for BP nanolayers allowing excellent control of BP interaction with light.[22]

Here, we report the first experimental near-field optical nanoscopic investigation of BP at mid infrared frequencies. We have observed near-field amplitude patterns, which allow direct imaging of the gap field at the tip-sample junction via bright fringes formed at BP edges outside of the BP boundaries. In comparison with full electrodynamic modeling, we attribute them to a high surface polarizability consistent with surface-metallic, plasmonic behavior at mid-infrared (mid-IR) frequencies. Note that such surface charging is a well-known phenomenon for conventional 3D semiconductors[23] but has never been previously reported for BP or other 2D materials.

Near-field spectroscopic study has determined that this plasmonic behavior of BP exists up to the maximum excitation frequency of $\omega_m \approx 1176 \, cm^{-1}$, which yields an estimate of the bulk plasma frequency of the surface metal layer as $\hbar\omega_p \sim 0.4 \, eV$. This corresponds to carrier density $n \approx 9.3 \times 10^{19} \, cm^{-3}$ allowing one to estimate the thickness of the surface metal layer from the corresponding Thomas-Fermi screening radius $r_{TF} = \hbar\varepsilon_{BP}^{1/2}(2e)^{-1}m_e^{-1/2}(3n/\pi)^{-1/6} \sim 1 \, nm$, where $\varepsilon_{BP}$ is bulk permittivity of BP, and $e$ and $m_e$ are electron charge and mass. Note that calculation with classical Debye-Hückel screening length $r_{DH} = \left(\varepsilon_{BP}k_BT\right)^{1/2}\left(4\pi ne^2\right)^{-1/2} \sim 1 \, nm$, where $k_B$ is Boltzmann constant, and $T$ is temperature, yields the same estimate due the fact the Fermi energy of this surface carrier gas is on the order of $k_BT$. Fit of the electrodynamic modeling to the experiment suggests that dielectric permittivity of the surface metallic layer is $\varepsilon \approx -5 + i0.5$ corresponding to Drude behavior with three-dimensional electron density $n \approx 1.2 \times 10^{20} \, cm^{-3}$, which is consistent with that found from the spectroscopic data.

Using spectroscopic imaging we have also shown that similar behavior takes place in other two-dimensional semiconductors such as TMDCs and topological insulators but not in insulators such as boron nitride. Our results suggest that this new phenomenon is related to surface charging of semiconductor nanofilms that are highly polarizable.

Near-field optical images were acquired using a commercial s-SNOM system (neaspec.com) represented schematically in Fig. 1a. A linearly polarized mid infrared quantum cascade laser (Daylight Solutions) is focused on the tip–sample interface at an angle of $45^0$ to the sample surface, and the scattered field is detected by phase-modulation interferometry. Topography (Fig. 1b) and third harmonic near-field amplitude images of a wedge-shaped uncoated BP exfoliated flake on a Si/SiO$_2$ substrate are shown at two frequencies (Figs. 1d,e ). Near-field amplitude images displayed in Figs. 1d,e show bright contrast compared to the substrate. The amplitude image taken at $\omega$ =934.6 cm$^{-1}$ (Fig. 1d) shows a bright fringe surrounding the wedge separated by a dark contrast from the inner bright surface of the structure. Such an edge fringe is missing in the near-field amplitude image taken at $\omega$ =1818.2 cm$^{-1}$ (Fig. 1e), suggesting a strong frequency dependence of the fringe formation. In fact, the fringe disappears at a critical frequency $\omega_p \approx 1176$ cm$^{-1}$ yielding an estimate of electron density given in the previous paragraph, $n \approx 9.3 \times 10^{19} \, cm^{-3}$.

The observed fringe in Fig. 1d is right outside the geometric end of BP structure as shown by the broken straight line drawn across the line profile plots of the three images shown in Fig. 1b. Similar outside bright fringes followed by a dark contrast fringe are seen in Fig. 1g on BP coated with 1 nm $Al_2O_3$ and on uncoated BP of the identical thickness (Fig. 1f). We have observed such edge fringes in BP on other substrates such as $Al_2O_3$ and GaAs, as well as on BP itself.

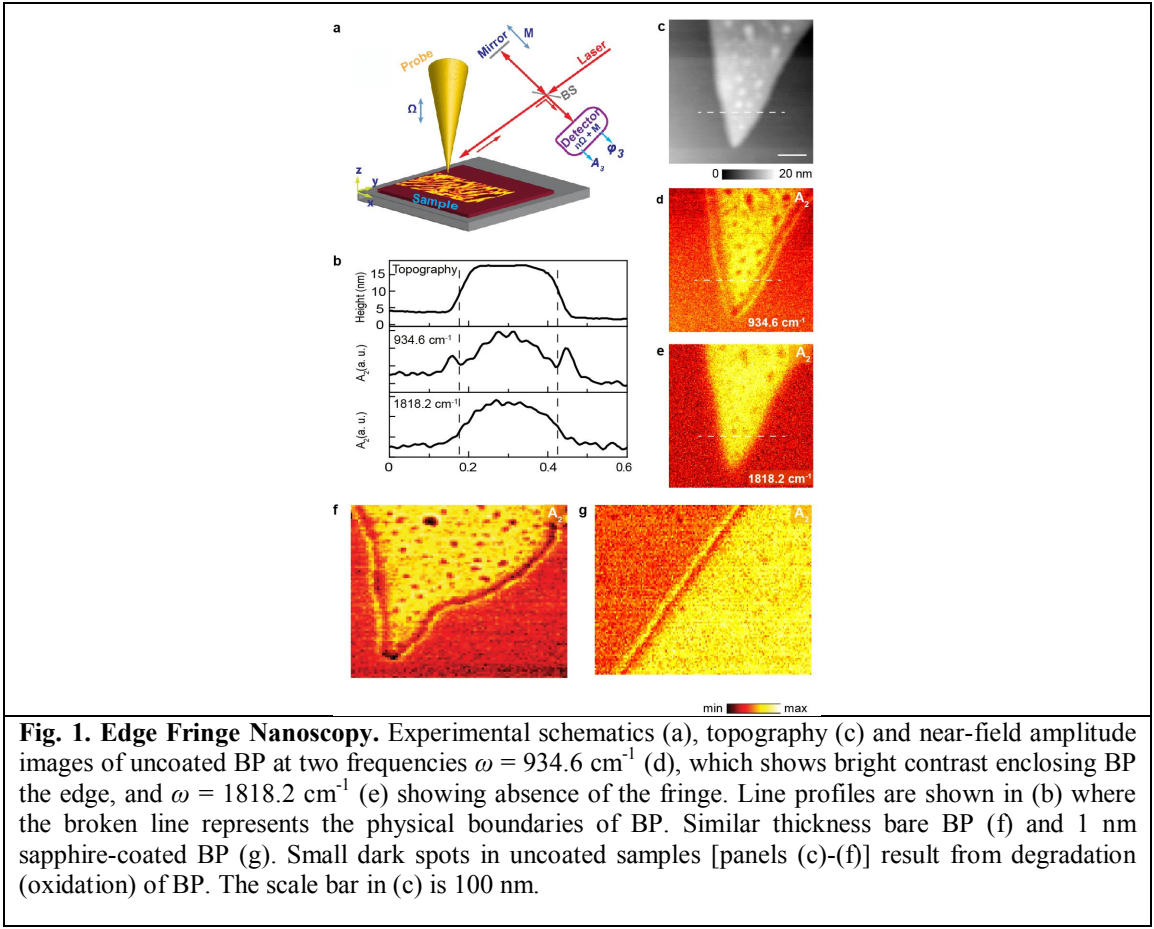

**Fig. 1. Edge Fringe Nanoscopy.** Experimental schematics (a), topography (c) and near-field amplitude images of uncoated BP at two frequencies $\omega = 934.6$ cm$^{-1}$ (d), which shows bright contrast enclosing BP the edge, and $\omega = 1818.2$ cm$^{-1}$ (e) showing absence of the fringe. Line profiles are shown in (b) where the broken line represents the physical boundaries of BP. Similar thickness bare BP (f) and 1 nm sapphire-coated BP (g). Small dark spots in uncoated samples [panels (c)-(f)] result from degradation (oxidation) of BP. The scale bar in (c) is 100 nm.

The near-field edge fringe formation is elucidated using theoretical simulation that takes into account the full structure of the probe tip, BP sample and $SiO_2$ substrate (Fig. 2). With the electron density $n \approx 1 \times 10^{20}$ cm$^{-3}$, as we have already mentioned in the introductory part of his article, the Thomas-Fermi and Debye-Hückel screening radii are ~1 nm suggesting that there is a thin metallic layer at the surface of the BP. This metallic nanofilm is likely originated from band bending caused by the surface charges as known for conventional 3D semiconductors.[23]

Assuming such a 1-nm metallic film at the BP surface, we performed full electrodynamic modeling of the near fields using time domain software from both Lumerical and CST Microwave Studio, obtaining consistent results. As Figs. 2(b)-(d) shows, when the s-SNOM tip is close to the edge of the sample, a gap "hot-spot" field is produced between the tip and the sample edge, reminding hot spots in metal nanoplasmonics.[24] The field in the hot spot is dependent on the permittivity of the sample surface – see Fig. 2(a). A bright fringe followed by a dark contrast that precedes a bright area on the sample surface is achieved when a metallic layer ($\text{Re}\varepsilon < 0$) is present at the surface, which is a requirement for plasmonicity.[25] The outside-fringe contrast increases with increasing $-\text{Re}\varepsilon$, consistent with plasmonic high polarizability of the sample surface. The best simulation of the experimental data are obtained for $\varepsilon = -5 + i0.5$. From this value, employing Drude formula $\varepsilon = \varepsilon_0 - \omega_p^2 / \omega(\omega + i\gamma)$, where $\varepsilon_0$ is the background permittivity and $\omega_p$ is the bulk plasma frequency, assuming $\varepsilon_0 = 6$ as typical for a narrow-band semiconductor, we determine $\omega_p \approx 0.4 \text{ eV}$, which is a value reasonable for highly-doped, conducting semiconductors.[26] Spatially, the bright fringe is detected to be outside of the BP edge by a distance approximately equal to the s-SNOM tip apex radius of curvature, which is ~20 nm. This is in a good qualitative agreement with experimental near-field images of Figs. 1(d)-(g).

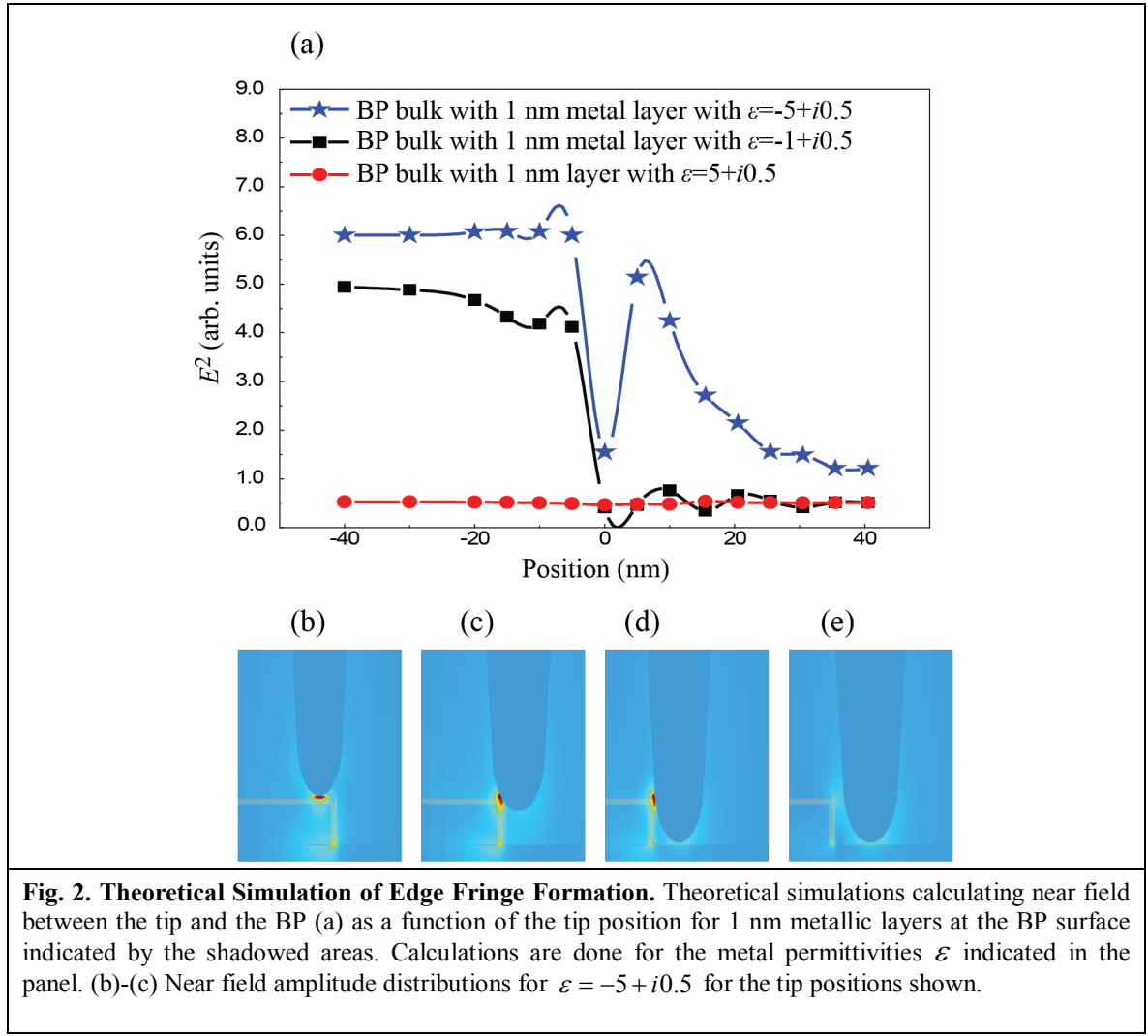

**Fig. 2. Theoretical Simulation of Edge Fringe Formation.** Theoretical simulations calculating near field between the tip and the BP (a) as a function of the tip position for 1 nm metallic layers at the BP surface indicated by the shadowed areas. Calculations are done for the metal permittivities $\varepsilon$ indicated in the panel. (b)-(c) Near field amplitude distributions for $\varepsilon = -5 + i0.5$ for the tip positions shown.

To better understand the frequency dependence of the observed edge fringe contrast in Fig. 1 and Fig. 2, we performed spectroscopic imaging of BP nanolayer at a series of frequencies. The thickness of the BP sample is $h \sim 20$ nm (topography shown in Fig. 3a) with a ~1 nm of $Al_2O_3$ coating deposited to slow oxidation degradation. Figure 3b shows a plot of the normalized fourth harmonic near-field amplitude for BP exfoliated on the SiO$_2$ substrate as a function of excitation frequency. As shown in Fig. 3b, for the frequency range between 1000 cm$^{-1}$ to 1200 cm$^{-1}$, the normalized amplitude signal is less than 1 (dark contrast of BP with respect to the substrate), while for all other frequencies the normalized amplitude value is greater than 1 (bright contrast of BP with respect to the substrate). This plot indicates highly frequency-dependent near-field contrast formation of the edge fringe for BP, observed for the first time.

In Figs. 3c-e, we show optical near-field amplitude images at three selected frequencies taken from the normalized amplitude plot Fig. 3b. The images faithfully represent the plot showing BP contrast darker than substrate for $\omega$ = 1176.5 cm$^{-1}$ (Fig. 3c) and $\omega$ = 1052.6 cm$^{-1}$ (Fig. 3d), which implies a smaller real part of the BP permittivity compared to SiO$_2$. For $\omega$ = 934.6 cm$^{-1}$ (Fig. 3e), the BP surface is brighter than substrate showing larger $|\text{Re}\,\varepsilon|$ of BP compared to that of the SiO$_2$ substrate. Moreover, the fringe around the nanostructured BP is clearly visible at $\omega$ = 934.6 cm$^{-1}$ (Fig. 3e) and turns off at $\omega$ = 1176.5 cm$^{-1}$ (Fig. 3c). At $\omega$ = 1052.6 cm$^{-1}$ (Fig. 3d) the background contrast is low compared to the SiO$_2$ substrate due to strong phonon mode of SiO$_2$, yet the fringe around BP is discernible. Similar to Figs. 1d,e, the results shown in Fig. 3 are consistent with the theoretical prediction that the fringe occurrence is dependent on the permittivity of BP.

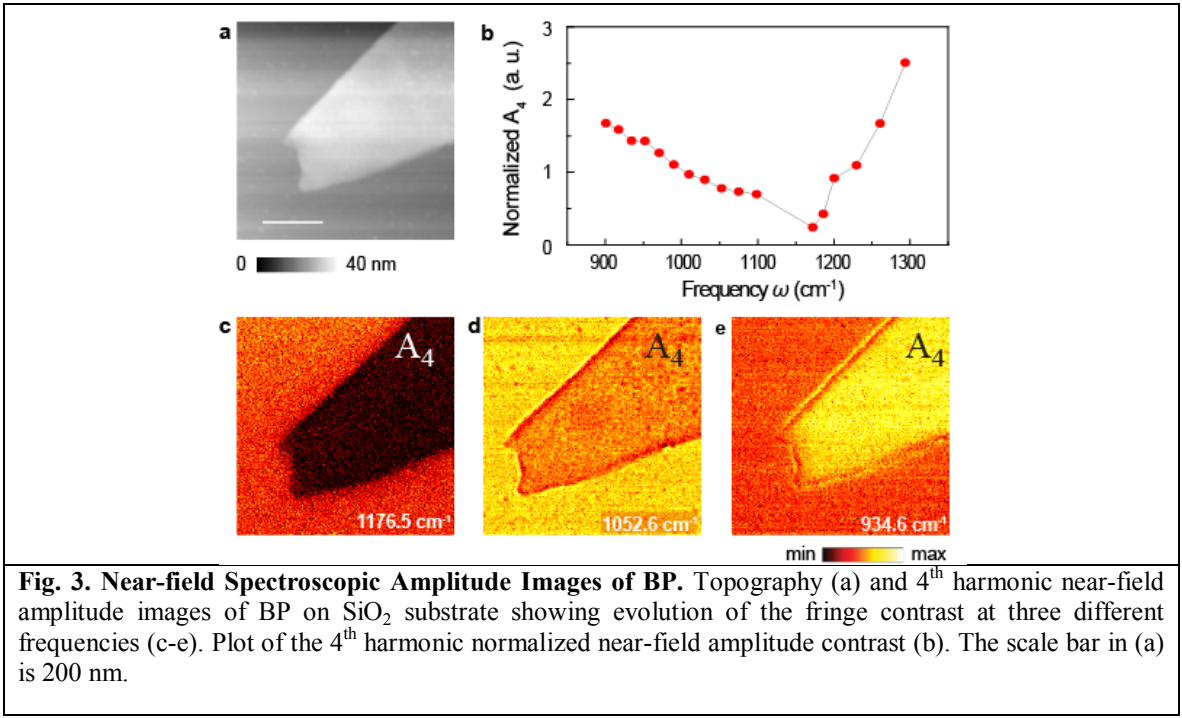

**Fig. 3. Near-field Spectroscopic Amplitude Images of BP.** Topography (a) and 4$^{th}$ harmonic near-field amplitude images of BP on SiO$_2$ substrate showing evolution of the fringe contrast at three different frequencies (c-e). Plot of the 4$^{th}$ harmonic normalized near-field amplitude contrast (b). The scale bar in (a) is 200 nm.

To test the universal nature of the observed outside-fringe formation, which we suggest to occur for sharp edge materials with surface-metallic behavior, we performed similar experiments using several 2D materials on identical SiO$_2$ substrate imaged at 934.5 cm$^{-1}$ where the strongest fringe contrast was observed for BP. As shown in Fig. 4, we find a similar, albeit weaker, edge fringe for nano-layered semiconductors MoS$_2$ (Fig. 4a) and Bi$_2$Te$_3$ (Fig. 4b) with a similar height ($h$~20 nm), which suggests electronic plasmonic behavior. However h-BN images show no fringes at $\omega = 934.6$ cm$^{-1}$ (Fig. 4d), a behavior expected for an insulator missing electronic plasmonic response. However, when tuned to a phonon-polariton (reststrahlen) spectral-range frequency, $\omega = 1562.5$ cm$^{-1}$ (Fig. 4c), h-BN shows the expected fringes inside the structure (Fig. 4c), in agreement with the recent publication.[27]

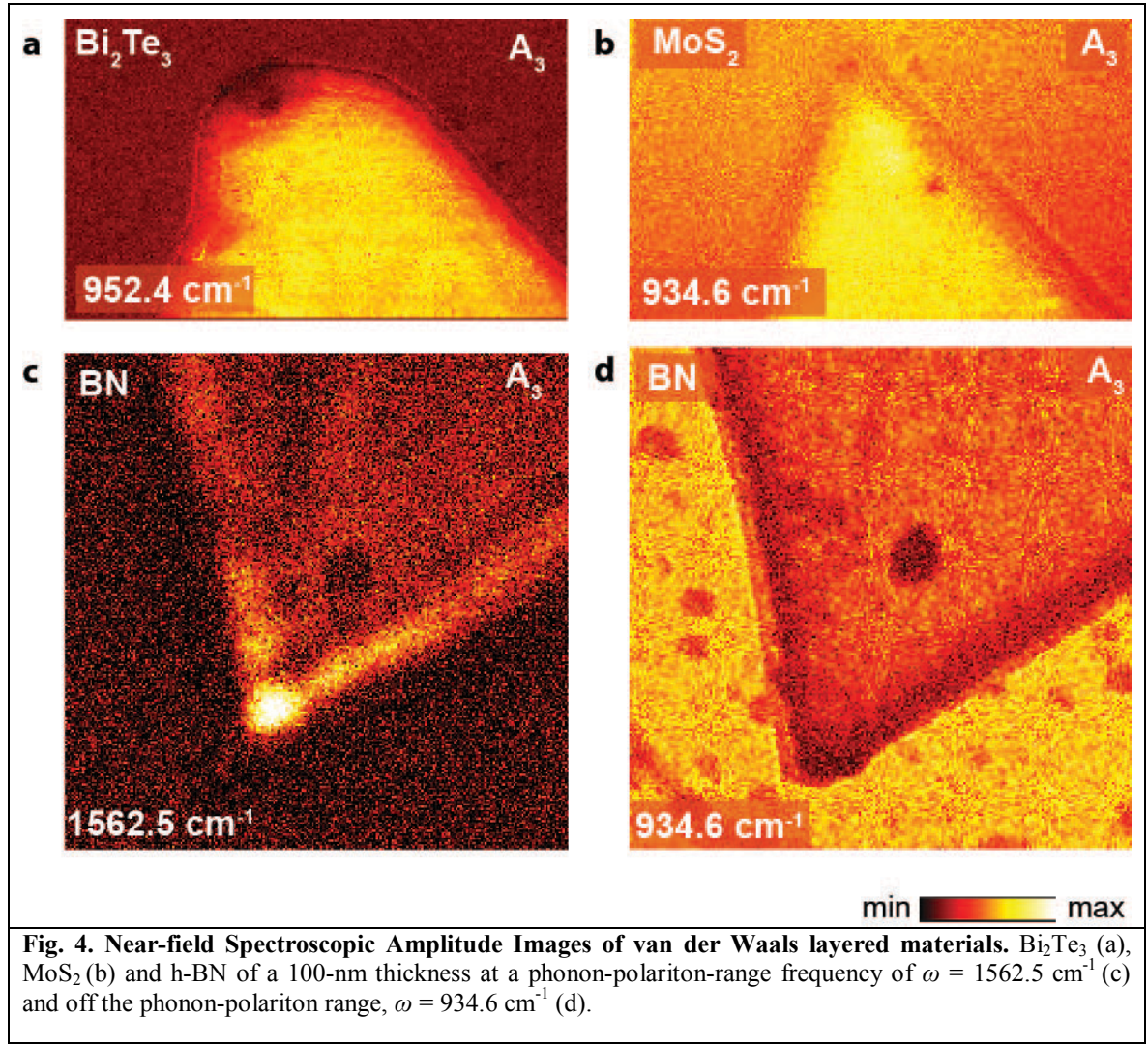

**Fig. 4. Near-field Spectroscopic Amplitude Images of van der Waals layered materials.** Bi$_2$Te$_3$ (a), MoS$_2$ (b) and h-BN of a 100-nm thickness at a phonon-polariton-range frequency of $\omega = 1562.5$ cm$^{-1}$ (c) and off the phonon-polariton range, $\omega = 934.6$ cm$^{-1}$ (d).

Non-encapsulated BP layers chemically degrade in ambient conditions.[28] We investigated the effect of degradation on the observed surface-metallic behavior by imaging a bare (unprotected) exfoliated BP sample left in the air to degrade for several weeks .We performed near-field optical imaging in amplitude and phase at two frequencies, where strong fringe is expected ($\omega = 934.6$ cm$^{-1}$) and at another, higher frequency ($\omega = 1190.5$ cm$^{-1}$). As shown in Fig. 5b, the bright edge fringe around the nanostructure is missing at $\omega = 934.6$ cm$^{-1}$ where it is expected for non-degraded sample. We attribute it to alteration of the surface properties due to degradation causing the surface to lose its metallic behavior. We note, however, that the surface near the edges of the samples degrades last as shown by the bright amplitude contrast at $\omega = 934.6$ cm$^{-1}$ (Fig. 5b) which turns to a darker contrast at $\omega$ = 1190.5 cm$^{-1}$ (Fig. 5c). The phase images show a large phase contrast at $\omega$ = 1190.5 cm$^{-1}$ (Fig. 5e) compared to that at $\omega$ = 934.6 cm$^{-1}$ (Fig. 5d). Based on the spectroscopic normalized amplitude and phase plots of degraded and non-degraded BP, we identify the bright portion shown in Fig. 5b as relatively intact BP and the dark large region as degraded BP. For detailed analysis of amplitude and phase spectra of BP and degraded BP.

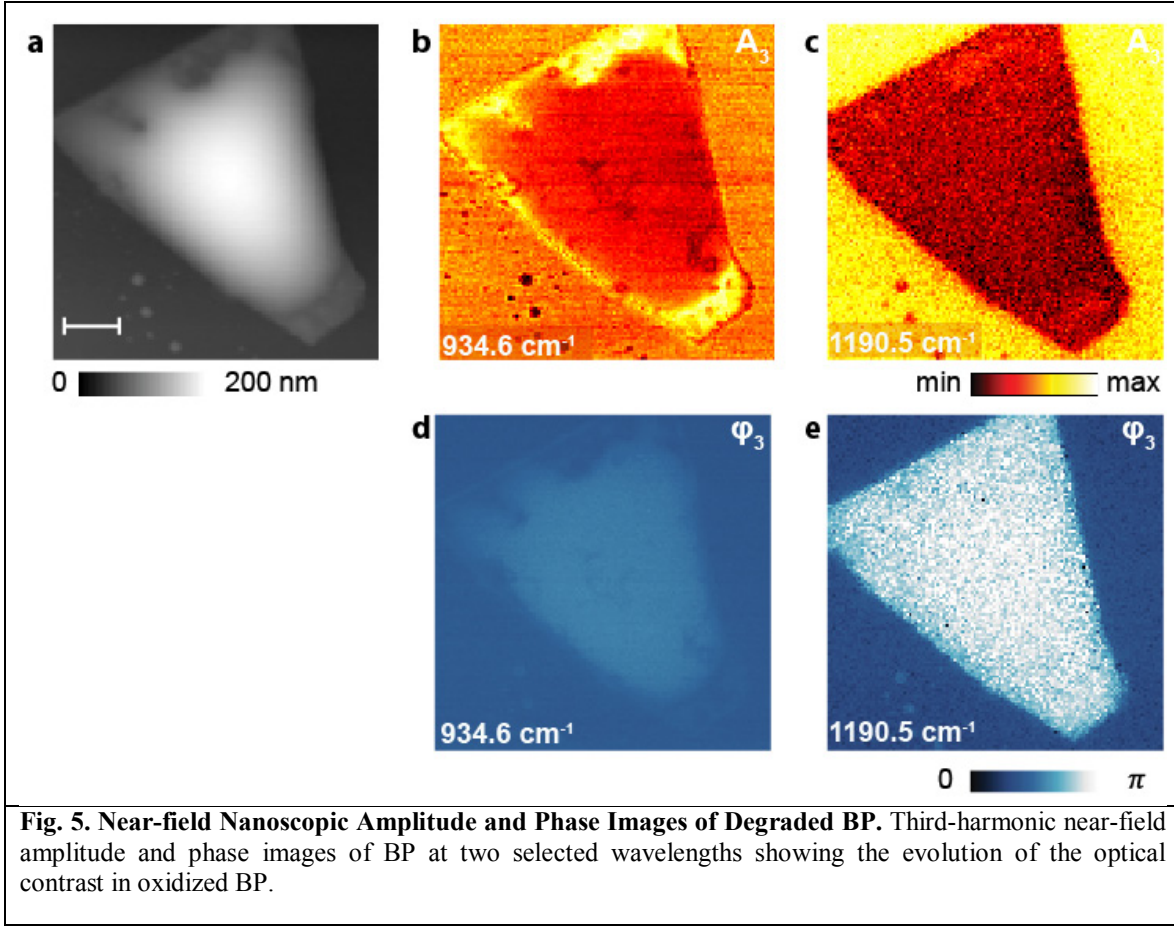

**Fig. 5. Near-field Nanoscopic Amplitude and Phase Images of Degraded BP.** Third-harmonic near-field amplitude and phase images of BP at two selected wavelengths showing the evolution of the optical contrast in oxidized BP.

In summary, we have shown the first experimental evidence that BP exhibits highly-polarizable surface-metallic behavior in the mid infrared frequency region as revealed via near-field edge fringe formation. This behavior exists only for frequencies below a threshold plasmon frequency found to be $\omega_p \approx 1176$ cm$^{-1}$, which gives an estimate for the electron concentration in the surface layer as $n \sim 10^{20}$ cm$^{-3}$, which is high enough to result in metallic properties of the surface layers. The thickness of the surface layer can be estimated from Thomas-Fermi or Debye-Hückel screening radius as $\sim 1$ nm. This metallic behavior is likely to result from surface charges due to dangling bonds, a phenomenon well known for 3D semiconductors but observed here for the first time for 2D materials. We have observed a similar surface metallic behavior is for other 2D (layered) semiconductors but not for dielectric layered material BN. We envision that such metallic properties and strong nanoplasmonic interaction of mid-infrared light with BP, which can potentially be gate-tunable, will open up diverse applications in nanooptics, electronics, and optoelectronics.

## Methods

**Near-field microscopy.** The microscope is a commercial s-SNOM system (neaspec.com). A probing linearly p-polarized QCL laser is focused on the tip–sample interface at an angle of $45^0$ to the sample surface. The scattered field is acquired using a phase modulation (pseudoheterodyne) interferometry. The background signal is suppressed by vertical tip oscillations at the mechanical resonance frequency of the cantilever ($f_0 \sim$ 285 kHz) and demodulation of the detector signal at higher harmonics $nf_0$ (commonly $n$=2,3,4) of the tip resonance frequency. The combined scattered field from the tip and the reference beam pass through a linear polarizer, which further selects the p/p polarization of the measured signal for analysis.

**Sample fabrication.** Black phosphorous flakes were exfoliated using mechanical exfoliation onto oxidized silicon wafers. Thin coatings of $Al_2O_3$ were deposited using atomic layer deposition at $220^oC$ with trimethyl aluminum as the Al source and water vapor as the oxygen source

**Acknowledgments**

YA acknowledges the major support for this work from DOD (U.S. Army Research Office) Grant No. W911NF-12-1-0076. Work of MIS was supported by Grant No. DE-SC0007043 from the Materials Sciences and Engineering Division of the Office of the Basic Energy Sciences, Office of Science, U.S. Department of Energy. The work at USC was supported by NSF grant CBET-1402906.

**Author contributions**

YA and SG performed the experiments. MIS, YA, SBC, and HW discussed the results. LZ, SBC, and HW fabricated the samples. VB, MHJ, and MIS performed theoretical work and carried out numerical simulations. YA and MIS wrote the text of the paper.

**Competing financial interests**

The authors declare no competing financial interests.